\documentclass[english, 12pt]{article}
\usepackage[T1]{fontenc}
\usepackage[utf8]{inputenc}
\usepackage[top=1.25in, bottom=1.25in, left=1.25in, right=1.25in]{geometry}
\usepackage[pdftex]{graphicx}
\usepackage{verbatim}
\usepackage[width=0.8\textwidth]{caption}
\usepackage{amsmath,amssymb,amsopn,dsfont,mathrsfs, bm}
\usepackage[english]{babel}
\usepackage{lmodern}
\usepackage{flafter, float,placeins,longtable,array}
\usepackage{natbib}
\usepackage{euscript}
\usepackage{color}
\usepackage{lscape,subfig,threeparttable}

\usepackage{pdflscape}
\usepackage{afterpage}

\usepackage{multirow}
\usepackage{enumitem}
\usepackage{booktabs}
\usepackage{stmaryrd}
\usepackage[hidelinks]{hyperref}
\definecolor{navyblue}{rgb}{0.0, 0.0, 0.5}
\hypersetup{colorlinks=true, citecolor=navyblue, linkcolor=black, urlcolor=black}
\makeatletter
\newtheorem{thm}{Theorem}[section]

\newtheorem{hyp}{Assumption}

\numberwithin{equation}{section}

\newcommand{\R}{\mathbb R}

\renewcommand{\triangleq}{:=}

\renewcommand{\section}{\@startsection{section}{2}{0mm}{-1.5\baselineskip}{1\baselineskip}{\normalfont\Large\bfseries}}
\renewcommand{\subsection}{\@startsection{subsection}{2}{0mm}{-1.2\baselineskip}{1\baselineskip}{\normalfont\normalsize\bfseries}}
\renewcommand{\subsubsection}{\@startsection{subsubsection}{3}{0mm}{-0.8\baselineskip}{0.4\baselineskip}{\normalfont\normalsize\itshape}}

\date{}

\begin{document}
	\title{On the Existence of Conditional Maximum Likelihood Estimates of the Binary Logit Model with Fixed Effects\thanks{I would like to thank Xavier D'Haultf\oe uille and Louis-Daniel Pape for their comments. All remaining errors are mine.}}
	\author{Martin Mugnier\thanks{CREST, \href{mailto:martin.mugnier@ensae.fr.}{martin.mugnier@ensae.fr}.}}

\maketitle

\begin{abstract}
By exploiting \cite{McFa73}'s results on conditional logit estimation, we show that there exists a one-to-one mapping between existence and uniqueness of conditional maximum likelihood estimates of the binary logit model with fixed effects and the  configuration of data points. Our results extend those in \cite{Anderson_1984} for the cross-sectional case and can be used to build a simple algorithm that detects spurious estimates in finite samples. As an illustration, we exhibit an artificial dataset for which  the STATA's command \texttt{clogit} returns spurious estimates.
 
\end{abstract}
	
	\newpage

\section{Introduction}\label{sec:intro}
Suppose there are $i=1,\dotsc,n$ individuals who are observed making a choice from $j \in \{0,1\}$ alternatives, over $t=1, \dotsc, T$ choice situations. Individual $i$'s sequence of choices is $Y_{i}:=(Y_{i1}, \dotsc, Y_{iT})$ with elements $Y_{it}=\mathds 1\{i \text{ chooses } 1 \text{ in } t\}$ that indicate $i$'s choice in choice situation $t$. Suppose in addition to observe individual-specific covariates $X_{it}\in \R^p$ for individual $i$ in choice situation $t$. Individual $i$'s sequence of covariates is then $X_i:=(X_{i1}', \dotsc, X_{iT}')'$. Let $\mathbf{Z}_n:=(Y_i,X_i)_{1\leq i\leq n}$ denote the sample data. Provided it exists and is unique, the conditional maximum likelihood estimator (that we denote by $\widehat{\beta}_n^{\mathrm{CMLE}}$) verifies
\begin{equation}\label{eq:estimator_cmle}
    \widehat{\beta}_n^{\mathrm{CMLE}}:= \arg \max_{\beta \in \R^p} \,  \log L(\beta; \mathbf{Z}_n)\,,
\end{equation}
where 
\begin{equation*}
    \log L(\beta; \mathbf{Z}_n) := \sum_{i=1}^N \log \frac{\exp\left(\sum_{t=1}^T Y_{it}X_{it}'\beta\right)}{\sum_{d_i:\sum_{t=1}^T(d_{it}-Y_{it})=0} \exp\left(\sum_{t=1}^Td_{it}X_{it}'\beta\right)}\,.
\end{equation*}

\medskip
As \cite{Andersen_1970} showed \citep[see also][for an earlier reference]{rasch1961}, $\widehat{\beta}_n^{\mathrm{CMLE}}$ is consistent for the binary logit model with fixed effects. However, it may be the case that $\widehat{\beta}_n^{\mathrm{CMLE}}$ does not exist in finite samples (the maximum in \eqref{eq:estimator_cmle} is not unique, or the conditional log-likelihood does not have a maximum). While \cite{Anderson_1984} established necessary and sufficient existence conditions for maximum likelihood estimates in cross-sectional logistic regression models,\footnote{\cite{Anderson_1984} show that, under a rank condition on the design matrix, maximum likelihood estimates exist if and only if there is no binary linear classifier that perfectly predicts outcome from covariates for all data points lying outside the decision frontier.} such conditions for $\widehat{\beta}_n^{\mathrm{CMLE}}$ are lacking. Yet, these are of interest for at least two reasons. First, practitioners often use off-the-shelf programs with built-in nonlinear solvers which do not systematically detect, tag, or even report situations where estimates do {\em not} exist \citep[see, e.g.,][]{Cullough_2003}. Specifically, we show below with an artificial dataset that STATA's \texttt{clogit} stops after some iterations and returns spurious results, which further illustrates that the problem has not been widely reckognized.\footnote{Noteworthy, the ``\texttt{clogit}'' section in the Stata User's Guide does not discuss the existence problem at all.} Second, it is easily seen that $\widehat{\beta}_n^{\mathrm{CMLE}}$ does not take advantage of all the variation available in the data. This mechanically increases the probability of existence failure as we shall see. In this paper, we extend the data separation rules established in \cite{Anderson_1984} for cross-sectional logistic models. Our results hold under a rank condition which is similar to \citet[p. 2]{Anderson_1984}' full rank assumption imposed on the matrix of covariates (see Assumption~\ref{as:full_rank} below).

\medskip
In Section~\ref{sec:existence_thm}, we derive our main existence theorem and provide a simple algorithm to detect existence failures. In Section~\ref{sec:instances}, we exhibit an artificial dataset for which existence fails but \texttt{clogit} does not detect such failure and returns spurious estimates.

\section{Necessary and Sufficient Existence Conditions}\label{sec:existence_thm}
Following the terminology employed in \cite{McFa73}, the sample $\mathbf{Z}_n$ results from a choice experiment composed of $n$ distinct trials $(X_i, B_i)$, where $B_i$ is the alternative set defined as
\begin{equation*}
B_i\triangleq \left\{d:=(d_1, \dotsc,d_T) \in \{0,1\}^T: \sum_{t=1}^T(d_t - Y_{it})=0\right\}\,.
\end{equation*}
The alternative set contains $r_{ni}:={T \choose \sum_{t=1}^TY_{it}}$ alternatives $d_i^j$, indexed by $j=1, \dotsc,r_{ni}$, and with vectors of attributes $\sum_{t=1}^Td_{it}^jX_{it}$. Note that the number of alternatives, $r_{ni}$, differs from one individual to another. Let us define $r_n:=\sum_{i=1}^Nr_{ni}$. By rewriting Axiom 5 in \citet[p. 116]{McFa73} to fit our framework, we shall make the following assumption.
\begin{hyp}\label{as:full_rank}
For all $\beta \in \R^p$, the $r_n \times p$ matrix whose rows are 
$$\left(\sum_{t=1}^Td_{it}^jX_{it} - \sum_{s=1}^{r_{ni}}\frac{\sum_{t=1}^Td_{it}^sX_{it}\exp(\sum_{t=1}^Td_{it}^sX_{it}'\beta)}{\sum_{\ell=1}^{r_{ni}}\exp(\sum_{t=1}^Td_{it}^\ell X_{it}'\beta)}\right)'$$
for $j=1, \dotsc, r_{ni}$ and $i=1, \dotsc, N$ is of rank $p$.
\end{hyp}
Assumption~\ref{as:full_rank} holds when the data vary sufficiently across periods. A necessary condition is $r_n \geq p + N$. This is likely to hold in practice if $p$ has a reasonable size. If $N\geq p$, it will hold generally since $r_{ni}\geq 2$, but it may also hold for $N<p$ if $T$ is large. The following condition is an adaptation of Axiom 6 in \citet[p. 116]{McFa73}.
\begin{hyp}\label{as:quasi_comp_gen}
It does not exist $\beta^\ast \in \R^{p}\backslash\{0\}$ satisfying
$$\sum_{t=1}^T(d^j_{it} - Y_{it})X_{it}'\beta^\ast \geq 0$$
for $j = 1, \dotsc,r_{ni}$ and $i = 1, \dotsc, N$.
\end{hyp}
Assumption~\ref{as:quasi_comp_gen} is reminiscent of the separation and quasi-complete separation relationships (4) and (6) in \cite{Anderson_1984}. However, it is specific to the panel data structure considered here. 
\begin{thm}\label{thm:general_case}
Suppose Assumption \ref{as:full_rank} holds. Then, Assumption \ref{as:quasi_comp_gen} is necessary and sufficient for the existence of a finite and unique $\widehat{\beta}_n^{\mathrm{CMLE}}$. 
\end{thm}
Theorem~\ref{thm:general_case} gives a sufficient and necessary condition for existence and uniqueness of conditional maximum likelihood estimates that depends only on the configuration of data points. It follows from an application of Lemma 3 in \cite{McFa73}. We now turn to the problem of finding an automated procedure for detecting if Assumption~\ref{as:quasi_comp_gen} holds in practice. 
\begin{thm}\label{thm:auto_proc}
Suppose Assumption \ref{as:full_rank} holds. Then, Assumption \ref{as:quasi_comp_gen} holds if and only if the minimum in the following quadratic programming problem is zero:
$$\min_{u, \beta} u'u$$
subject to
\begin{equation*}
    u = \sum_{i=1}^N\sum_{j=1}^{r_{ni}}\beta_{ij}\sum_{t=1}^T(d^j_{it} - Y_{it})X_{it} \text{ and } \beta_{ij}\geq 1.
\end{equation*}
\end{thm}
Theorem~\ref{thm:auto_proc} follows from an application of Lemma 4 in \cite{McFa73}. Note that a \texttt{Python} module called \texttt{BinLogitCMLE} that implements the program given in Theorem~\ref{thm:auto_proc} before computing $\widehat{\beta}^{\mathrm{CMLE}}_n$ is made publicly available on the author's Github page.\footnote{\href{https://github.com/martinmugnier/BinLogitCMLE}{https://github.com/martinmugnier/BinLogitCMLE}.}

\section{An Example with Artificial Data}\label{sec:instances}
\begin{figure}[H]
    \caption{STATA's Logistic Regression}
    \label{fig:logit}
    \includegraphics[scale=0.8]{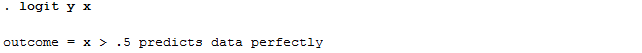}
\end{figure}

\begin{figure}[H]
    \caption{STATA's Conditional Logit with Fixed Effects}
    \label{fig:clogit}
    \includegraphics[scale=0.8]{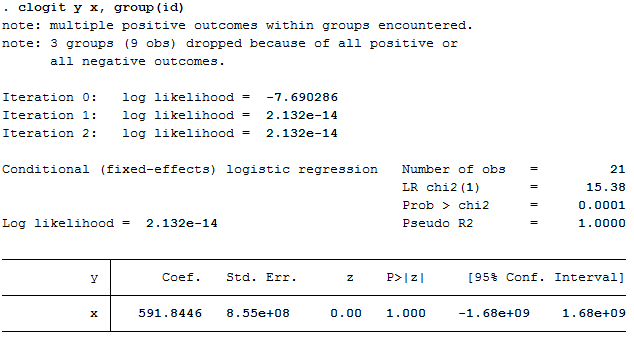}
\end{figure}

We generate an artificial dataset of $10$ individuals (with personal identifiers $\mathbf{id}$) who are observed at periods $\mathbf{t}\in\{1,2,3\}$. The choice variable is  $\mathbf{y}_{it} \in \{0,1\}$ and there is a unique regressor $\mathbf{x}_{it} \in [0,1]$. The data is displayed in Table~\ref{tab:synth_data}. 
\begin{table}[H]
	\centering
	\begin{threeparttable}
       \caption{Artificial Data}
		\label{tab:synth_data}
		\small{
			\begin{tabular}{c c c c c c c c c}
    \toprule
    $\mathbf{id}$ & $\mathbf{t}$ & $\mathbf{x}_{it}$ & $\mathbf{y}_{it}$ & & $\mathbf{id}$ & $\mathbf{t}$ & $\mathbf{x}_{it}$ & $\mathbf{y}_{it}$ \\
    \cmidrule{1-4}\cmidrule{6-9}
    1 &	1 &	0.48 & 0 && 6 & 1 & 0.33 & 0 \\
    1 &	2 &	0.50 & 0 && 6 & 2 & 0.62 & 1 \\
    1 &	3 &	0.42 & 0 && 6 &	3 &	0.48 & 0 \\
    2 &	1 &	0.52 & 1 && 7 &	1 &	0.001 & 0 \\
    2 &	2 &	0.46 & 0 && 7 &	2 &	0.87 & 1 \\
    2 &	3 &	0.60 & 1 && 7 &	3 &	0.85 & 1 \\ 
    3 &	1 &	0.57 &1  && 8 &	1 &	0.78 & 1 \\ 
    3 &	2 &	0.58 & 1 && 8 &	2 &	0.95 & 1 \\
    3 &	3 &	0.39 & 0 && 8 &	3 &	0.51 & 1 \\ 
    4 &	1 &	0.40 & 0 && 9 &	1 &	0.26 & 0 \\ 
    4 &	2 &	0.37 & 0 && 9 &	2 &	0.99 & 1 \\
    4 &	3 &	0.52 & 1 && 9 & 3 & 0.43 &  0 \\
    5 &	1 &	0.10 & 0 && 10 & 1 & 0.17 & 0 \\   
    5 &	2 &	0.41 & 0 && 10 & 2 & 0.22 & 0 \\
    5 &	3 &	0.25 & 0 && 10 & 3 & 1 & 1 \\
\bottomrule
			\end{tabular}
}
\end{threeparttable}
\end{table}
Note that $\mathbf{y}_{it}=1$ if and only if $\mathbf{x}_{it}>0.5$. Hence, the stacked data is separated in \cite{Anderson_1984}' sense. Actually, it is easy to check that the data also violate Assumption~\ref{as:quasi_comp_gen}. While the \texttt{logit} program detects the separation (see Figure~\ref{fig:logit}), \texttt{clogit} does not detect violation of Assumption~\ref{as:quasi_comp_gen} and returns spurious estimates after three iterations (see Figure~\ref{fig:clogit}).  Although the estimated standard error (resp. the log-likelihood function) is quite large (resp. almost zero), the user may miss the crucial fact that these results are informative only about the nonexistence of the estimate. We note that \texttt{BinLogitCMLE} detects the existence failure as expected.

\newpage
	\bibliography{biblio}
	
\newpage

\section{Appendix}

\end{document}